\documentclass[preprint, 5p, twocolumn]{elsarticle}   

\usepackage{mathrsfs,amsbsy,amssymb,latexsym,amsfonts,amsmath}

\newcommand\Lap{{\bigtriangleup}}
\newcommand\CH{\mathcal H}

\begin{document}


\title{CMB Fluctuations and String Compactification
Scales}
 
\author[oiqp]{Yoshinobu Habara}
\ead{habara@yukawa.kyoto-u.ac.jp}
\author[kyoto]{Hikaru Kawai}
\ead{hkawai@gauge.scphys.kyoto-u.ac.jp}
\author[oiqp]{Masao Ninomiya}
\ead{ninomiya@yukawa.kyoto-u.ac.jp}
\author[kek]{Yasuhiro Sekino}
\ead{sekino@post.kek.jp}
\address[oiqp]{Okayama Institute for Quantum Physics, 1-9-1 Kyoyama, 
Okayama 700-0015, Japan}
\address[kyoto]{Department of Physics, Kyoto University, 
Kyoto 606-8502, Japan}
\address[kek]{Theory Center, Institute for Particle and Nuclear Studies, 
KEK, Tsukuba 305-0801, Japan}


\begin{abstract}
We propose a mechanism for the generation of temperature fluctuations 
of cosmic microwave background. We consider a large number of fields,
such as Kaluza-Klein modes and string excitations. Each field
contributes to the gravitational potential by a small amount, 
but an observable level of temperature fluctuations is achieved
by summing up the contribution of typically of order $10^{14}$ fields.
Tensor fluctuations are hardly affected by these fields. Our mechanism 
is based on purely quantum effects of the fields which are classically 
at rest, and is different from the one in slow-roll inflation.
Using the observed data, we find constraints on the parameters
of this model, such as the size of the extra dimensions and
the string scale. Our model predicts a particular pattern of
non-gaussianity with a small magnitude.
\end{abstract}


\maketitle

\paragraph*{Introduction.---}

Observation of cosmic microwave background (CMB)~\cite{WMAP} provides
a great opportunity for testing fundamental theories. It is generally
believed that the fluctuations of CMB originate from
quantum fluctuations generated at energy  much higher than
present accelerators can achieve. 

It is hoped that string theory gives predictions for observational
cosmology. But at the current stage of its development, string
theory is defined only in limited classes of background spacetimes.
The analysis of cosmology has been limited to those based on low energy 
effective field theory, and it has been difficult to make concrete predictions.
(For attempts to formulate cosmology beyond the level of
effective field theory, see e.g.\ \cite{HFKN} and \cite{FSSY}.)

In this Letter we will focus on an aspect of fundamental theories, 
namely the presence of a large number of fields. These could be 
Kaluza-Klein (KK) modes from the 
compactification of extra dimensions or excited 
states of strings. Our analysis is mostly based on effective 
field theory, but we will also use a general property of
perturbative string theory as an input. 

It would be possible that the size of extra dimensions $L$ is
large in unit of Hubble scale of inflation $H^{-1}$. 
We already know a hierarchy of scales between the Planck scale 
$m_{pl}$ and $H$: The fact that tensor perturbation
(B-mode polarization of CMB) has not been observed implies
$H$ is at least 4 orders of magnitude smaller than $m_{pl}$~\cite{WMAP}. 
The question is whether string scale $m_{s}$ is close to $m_{pl}$
or $H$. We will see below that with $m_{s}$ slightly smaller than
$H\sim 10^{-4}m_{pl}$,
$L$ can be as large as $L\sim 10^{8}H^{-1}$. 
In such a case, there will be many KK modes with $m\lesssim H$. 
We will show that the quantum effects of these fields lead to
interesting observable effects.

For concreteness, we shall consider a collection of $N$ free 
fields $\phi_{A}$ ($A=1,\ldots, N$) with mass $m_{A}$, which
are classically at the bottom of the potential
$\phi_{A}=0$. We will not consider inflaton
field, and take the background to be pure de Sitter,
and find the temperature fluctuations generated 
during inflation. At the end of this Letter, we 
make comments on the effect of time dependence of
Hubble, e.g.\ near the end of inflation.

In this Letter we point out that temperature fluctuations 
$\delta T/T$ can be
generated by a mechanism different from the standard one 
based on slow-roll inflation.
In the latter case, the classical value of a scalar field (inflaton)
provides a preferred time slicing, and its fluctuation is interpreted
as the difference of duration of inflation at different points in
space, which results in the density (or temperature) anisotropy. 
In our case, the fields $\phi_{A}$ are classically at rest, 
and the above argument does not apply. It has been assumed that
these fields do not contribute to $\delta T/T$. We will show that 
by keeping quadratic terms in $\phi_{A}$, the gravitational 
potential $\Phi$ is generated 
through quantum effects of these fields. Each field gives
small contribution, but when $N\gg 1$, this effect 
becomes important. The large ratio between scalar and tensor
contributions to $\delta T/T$ is usually assumed to be 
the consequence of the smallness of the slope of the inflaton
potential. In our approach, it is the consequence of a large number of fields.

In the following, we first review quantization in de Sitter space,
and recall that fields with $m<{3\over 2}H$ 
do not oscillate at super-horizon scales, because the ``friction''
due to the cosmic expansion overdumps the oscillation. 
These are the fields that we will focus on.  
We then study Einstein equations and solve for the gravitational
potential $\Phi$ 
in terms of the matter fields. Using this, we find 
the contribution to CMB fluctuations from KK modes.
We will make an argument based on string perturbation theory
that if string scale $m_{s}$ is lower than $H$,
the summation over the mass is effectively cut off at $m_{s}$.
We then use the observed data on the amplitude of $\delta T/T$ 
and the tensor-to-scalar ratio $r_{t/s}$ to
constrain the parameters, such as $m_{s}$ and $L$. 

There have been inflationary models involving a large number
of fields. In ``N-flation''~\cite{Nflation} (or ``assisted 
inflation''~\cite{Liddle}), many fields (such as axions)
classically roll down the potential, collectively  
producing an effect similar to chaotic inflation.
Our mechanism is different from this, since we
are not assuming classical motion. Also note that our
mechanism is different from the curvatons scenario~\cite{curvaton}. 
We demonstrate that curvature fluctuations are generated 
during inflations without such a late time mechanism.

\paragraph*{Quantization in de Sitter background.---}

We consider background de Sitter space
\begin{equation}
 ds^2=dt^2-a^2(t)d\vec{x}^2, \quad a(t)=H^{-1}e^{Ht},
\label{deSitter}
\end{equation}
and a collection of $N$ free massive scalar fields, 
\begin{equation}
S=\sum_{A=1}^{N}
\int d^{4}x \sqrt{-g}\left\{\partial_{\mu}\phi_{A}\partial^{\mu}\phi_{A}
-m_{A}^2\phi_{A}^2 \right\}.
\end{equation}
In the following we will often use the conformal time,
$\tau=\int dt/a(t)=-e^{-Ht}$ ($-\infty\le \tau\le 0$),
and the rescaled field $\chi_A=a\phi_A$ which has the
standard kinetic term. We will suppress the label
$A$ hereafter for brevity.

Fourier component of $\chi$ satisfies 
the equation of motion (prime denotes $\partial_{\tau}$),
\begin{equation}
\chi_{\vec{k}}^{\prime \prime}(\tau)+\left\{|\vec{k}|^2
+\left(H^{-2}m^2-2\right)\frac{1}{\tau^2}\right\}
\chi_{\vec{k}}(\tau)=0.
\label{scalareom}
\end{equation}
Quantization is done by setting 
\begin{equation}
\chi (\tau ,\vec{x})
=\int {d^{3}k\over (2\pi )^{3}}{1\over \sqrt{2|\vec{k}|}}
\left[u_{\vec{k}}(\tau )a_{\vec{k}}e^{i\vec{k}\cdot \vec{x}}
+u_{\vec{k}}^{\ast}(\tau )a^{\dagger}_{\vec{k}}e^{-i\vec{k}\cdot \vec{x}}\right]
\end{equation}
where $u_{\vec{k}}(\tau)$ is the solution of (\ref{scalareom}) 
which approaches
$u_{\vec{k}}(\tau)\to e^{-i|\vec{k}|\tau}$ at early time $\tau\to-\infty$. 
This condition ensures that the flat space result is recovered 
in the short distance limit. The solution is given by
$u_{\vec{k}}(\tau) =\sqrt{\pi\over 2}e^{i{\pi\over 2}(\nu+{1\over 2})}
\sqrt{-|\vec{k}|\tau} H_{\nu}^{(1)}(-|\vec{k}|\tau)$ with 
$\nu=\sqrt{{9\over 4}-m^{2}H^{-2}}$.  Asymptotic behavior 
at the late times (in the super-horizon $|\vec{k}|/a\ll H$ limit) 
is $u_{\vec{k}}\sim (-|\vec{k}|\tau)^{-\nu+{1\over 2}}$. Time dependence 
factorizes from space dependence, as is clear from the
fact that $|\vec{k}|$ dependence drops out from (\ref{scalareom}) 
in this limit. Fields with small mass, $mH^{-1}<{3\over 2}$, 
do not oscillate in time. We will take the $mH^{-1}\ll 1$
limit in the following formulas, since this is the case of
importance for our applications, as explained
below. 

The equal time two-point function of $\phi$ at late times is 
\begin{eqnarray}
\langle \phi(\tau, \vec{x} )\phi(\tau, \vec{x}' )\rangle
&=& {1\over a^{2}(\tau)}\int {d^{3}k\over (2\pi)^3}{1\over 2|k|}
|u_k(\tau)|^{2}e^{i\vec{k}(\vec{x}-\vec{x'})}\nonumber\\
&\sim& {H^{2}\over 4\pi^2 \beta}
(Ha|\vec{x}-\vec{x'}|)^{-\beta},
\end{eqnarray}
where we have taken the $m^{2}H^{-2}\ll 1$ limit, and defined
\begin{equation}
 \beta={2\over 3}m^{2}H^{-2}.
\end{equation}

\paragraph*{Einstein equations.---}
Quantum fluctuations of the fields $\phi$ induce gravitational
potential. Let us look at the (0, $i$) and ($i$, $j$) components
of the Einstein equation,
\begin{eqnarray}
&&(\Psi'+\CH\Phi),{}_{i} =4\pi G\delta T^{\rm (S)}_{0i},
\label{E0i}\\
&&\left[\Psi''+\CH (2\Psi+\Phi)'+(2\CH'+\CH^2)\Phi
+{\Lap\over 2}(\Phi-\Psi)\right]\delta_{ij}\nonumber\\
&&\qquad -{1\over 2}(\Phi-\Psi),{}_{ij}=4\pi G\delta T^{\rm (S)}_{ij}
\label{Eij}
\end{eqnarray}
where $\CH={a'\over a}=-{1\over \tau}$. The l.h.s is the 
Einstein tensor expanded to the first order in metric fluctuations;
$\Phi$ and $\Psi$ are the two gauge invariant combinations constructed 
from the scalar modes (See e.g.\ \cite{Mukhanov}). On the r.h.s.\
we have energy momentum tensor which is quadratic in $\phi$, 
\begin{equation}
 \delta T_{\mu\nu}=\sum\left\{ \partial_{\mu}\phi\partial_{\nu}\phi
-{1\over 2}g_{\mu\nu}(\partial^{\rho}\phi \partial_{\rho}\phi
-m^{2}\phi^2)\right\}.
\end{equation}
The sum is taken over the species of the fields. 
The superscript ${\rm (S)}$ denotes the scalar part
(the part which can be written as derivatives of a scalar),
e.g., 
\begin{equation}
\delta T^{\rm (S)}_{0i}=\partial_{i}\left({1\over \Lap}\partial_{k}
\delta T_{0k}\right)={1\over \Lap}\partial_{i}\partial_{k}
\left(\phi'\partial_{k}\phi\right)
\end{equation}

We shall use the Einstein equation to express
$\Phi$ and $\Psi$ in terms of the matter fields $\phi$.
We first find the difference $\Phi-\Psi$ from the $(i\neq j)$ component
of (\ref{Eij}), 
\begin{equation}
\Phi-\Psi=-8\pi G \sum s,\quad
s\equiv {3\over 2\bigtriangleup^2}\partial_{i}\partial_{j}
(\partial_{i}\phi\partial_{j}\phi-{\delta_{ij}\over 3}
\partial_{k}\phi\partial_{k}\phi)\nonumber
\end{equation}
and substitute it into (\ref{E0i}) to get
\begin{eqnarray}
\Phi'+\CH \Phi&=&4\pi G \sum \Big\{
-2s' +{1\over \Lap}\partial_{i}
(\phi' \partial_{i}\phi)\Big\}.
\label{E0i2}
\end{eqnarray}
At late times, the r.h.s. goes like $(-\tau)^{\beta-1}$, which 
implies $\Phi\sim (-\tau)^{\beta}$. The part containing $s$ 
can be dropped when $\beta\ll 1$, since this gives smaller
contribution in correlation functions than the second 
term\footnote{In correlation functions, ${\partial\over \partial x^{i}}
\langle\phi(x)\phi(x')\rangle$
gives a factor of order $\beta$ (and further differentiation only 
gives order 1 factors like $-(\beta+1)$). Correlators involving
the first term of (\ref{E0i2}) necessarily contain more of these
factors than the ones involving only the second term of (\ref{E0i2}).}. 
Using the fact that time dependence 
($\phi\sim (-\tau)^{\beta/2}$)
enters as a multiplicative factor, and 
$\partial_{i}(\phi\partial_{i}\phi)={1\over 2}\Lap \phi^2$,
we find
\begin{equation}
 \Phi=-\sum \pi G \beta \phi^2. 
\end{equation} 
at late times and $\beta\ll 1$. 
This  
satisfies all the components
of Einstein equations.
The connected part of two-point function becomes\footnote{The
expectation value (one-point function) of $\Phi$ is UV divergent, 
but the two-point function is not. As long as we are studying 
the connected part of two-point functions, we do not need 
renormalization.}
\begin{eqnarray}
 \langle \Phi(\tau, x)\Phi(\tau, x')\rangle
&=&\sum 2(\pi G\beta)^2 \langle \phi (\tau, x) \phi (\tau, x')\rangle^2
\nonumber\\
&=&\sum {H^{4}G^2\over 8\pi^{2}}(Ha|\vec{x}-\vec{x'}|)^{-2\beta}.
\label{PhiPhi}
\end{eqnarray}

Next we consider tensor fluctuations. 
The transverse-traceless (TT) mode of gravitons
($\nabla^{i}h_{ij}$=$h^{i}_{i}$=0) is sourced by the
TT part of energy-momentum tensor, 
\begin{equation}
 h''_{ij}+2\CH h'_{ij}-\Lap h_{ij}=8\pi G \delta T^{(T)}_{ij}.  
\label{tensoreq}
\end{equation}
The general solution to this equation is the sum
of 
the solution $h^{(0)}_{ij}$ for the homogeneous equation
and a particular solution $h^{(1)}_{ij}$
which depends on $\delta T^{(T)}_{ij}$. 
This means that the tensor fluctuation consists of the
usual gravitational wave $h^{(0)}_{ij}$ on top of the
piece $h^{(1)}_{ij}$ which is determined by
 $\delta T^{(T)}_{ij}$.
Since the homogeneous equation is equivalent 
to massless scalar equation, $h^{(0)}_{ij}$ scales 
logarithmically (in space and time), giving rise to
the scale invariant spectrum with the amplitude
$H/m_{pl}$.  On the other hand, $h^{(1)}_{ij}$ decays
in time ($h^{(1)}_{ij} \sim (-\tau)^{\beta+2}$,
since $\delta T^{(T)}_{ij}\sim (-\tau)^{\beta}$),
so the effect of $\delta T^{(T)}_{ij}$ is unimportant at late times.

\paragraph*{KK modes and string states.---}

Let us now assume that a part of the extra dimensions is
compactified on a manifold with the size $L\gg H^{-1}$, 
and find temperature fluctuations produced by the KK modes.
For simplicity, we assume $D$ dimensions 
are compactified on $T^{D}$ with the same periodicity 
$L$ in all directions, and the other dimensions are compactified 
on a space with the string scale size. KK modes from
$T^{D}$ have mass $m^2=\sum_{a=1}^{D} (2\pi n_{a})^2/L^2$.
Assuming the level is sufficiently dense, 
the number of states is given by
$S_{D-1}|n|^{D-1}d|n|=S_{D-1}(L/2\pi)^{D} m^{D-1}dm$,
where $S_{D-1}=2\pi^{D/2}/\Gamma(D/2)$ is the volume of
the $D-1$ dimensional unit sphere. 

In this Letter, we only consider scalar fields. There are also
vector and tensor KK modes, but as long as we are considering
quantum fluctuations around vanishing expectation values, they
behave similarly to massive scalars, and just contribute
extra multiplicities\footnote{We do not have problems of 
anisotropy, since vectors and tensors have zero expectation
values. Even if they aquire non-zero expectation values after 
inflation, the anisotropy will be suppressed as $1/\sqrt{N}$ when
there are $N$ independent fields. See \cite{vector} for discussion
in a closely related context.}.

The two point function of $\Phi$ (\ref{PhiPhi}) 
becomes
\begin{equation}
 \langle \Phi\Phi\rangle
=c_{D}L^{D}
 \left({H\over m_{p}}\right)^{4}
\int_{0}^{m_{c}}dm m^{D-1} 
(Ha|\vec{x}-\vec{x'}|)^{-2\beta},
 \end{equation}
where $c_{D}=S_{D-1}/(4(2\pi)^{D+2})$.
The upper limit $m_{c}$ of the integration should be 
$m_{c}\sim {3\over 2}H$ as long as we are working in 
Einstein gravity. However, if
string scale is less than Hubble scale, $m_{s}< H$,
string states should be taken into account. 
In this case, we expect that the sum over the mass is effectively 
cut off at $m_{c}\sim m_{s}$ for the following reason. Let us assume 
the two-point function of $\Phi$ comes from the one-loop diagram in 
string theory. String theory can be regarded as a field theory with 
infinitely many fields, except that one-loop amplitude effectively 
has UV cutoff due to modular invariance; Schwinger proper time is 
cut off at string scale. There is no physical meaning to time interval 
shorter than string scale, or oscillations much higher than string
scale. This means the internal states in the loop which has mass 
much larger than string scale do not have physical 
effect\footnote{This argument is based on string theory in flat 
background, and it is not clear whether this can be applied
to an arbitrarily curved background, but we believe this is a 
reasonable estimate. 
In fact, the precise value of this cutoff does not affect our final 
conclusion: The preferred value of the string scale, being about one 
order lower than Hubble scale, is not modified even if we take the 
upper limit to be $(3/2)H$ instead of $m_s$.}.


\paragraph*{CMB fluctuations.---}
Temperature fluctuation of CMB is related to $\Phi$ at recombination
(at redshift $z\sim 1100$) by ${\delta T/ T} =-\Phi/3$. 
The angle $\theta$ on the sky corresponds
to the distance $d_{r}=2R_{r}\sin(\theta/2)$ where $R_{r}\sim H_{0}^{-1}$
is the radius of the surface of last scattering. 
We are interested in the modes outside the horizon at recombination,
which correspond to $3^{o}\le \theta$ ($l\le 60$). 
For these modes, $\Phi$ will be frozen (remain constant at fixed 
comoving distance) after inflation. Thus we will look at the
correlator at distance 
${(a_{e}/a_{r})}d_{r}=2 R\sin(\theta/2)=a_{e}|\vec{x}-\vec{x'}|$ 
at the end of inflation, where
$a_{e}$ and $a_{r}$ are the scale factors at the end
of inflation and at recombination. 
The radius $R=(a_{e}/a_{r})R_{r}$ will depend on the scale of inflation.
We will assume $RH\sim 10^{29}\sim e^{67}$ in the following. 

The power spectrum $C_{l}$ is defined by
$\langle {\delta T\over T}(\theta) {\delta T\over T}(0)\rangle
=\sum_{l=1}^{\infty} (2l+1) C_{l}P_{l}(\cos\theta)$. To find it,
let us expand 
\begin{equation}
(Ha|\vec{x}-\vec{x'}|)^{-2\beta}
\sim (2RH)^{-2\beta}(1-2\beta\log(\sin(\theta/2)),
\nonumber
\end{equation}
and recall
that $-2\log(\sin(\theta/2))$ is $1/(l(l+1))$ in harmonic
space. The square amplitude $\delta^{2}_{T}\equiv l(l+1)C_{l}$ becomes
\begin{eqnarray}
\delta^{2}_{T}&=&{2\over 3}c_{D}{L^{D}\over H^{2}}
 \left({H\over m_{pl}}\right)^{4}
\int_{0}^{m_{s}}dm m^{D+1}(2RH)^{-{4\over 3}m^{2}H^{-2}}\nonumber\\
&=&{2\over 3}c_{D}\left({m_{s}\over H}\right)^{2}(Lm_{s})^{D} 
\left({H\over m_{pl}}\right)^{4}{\cal M}_{D}(\zeta_{0}),
\label{amplitude}
\end{eqnarray}
where we have taken the upper limit to be $m_{c}=m_{s}$.
This is a good approximation even when $m_{s}>H$, since the 
contribution from the region $m\gtrsim H/10$ will be
strongly suppressed due to the factor 
$(2RH)^{-{4\over 3}m^{2}H^{-2}}$ in any case. We have defined
\begin{equation}
 {\cal M}_{D}(\zeta)=\int_{0}^{1}dt e^{-\zeta t^2}t^{D+1}, \quad
\zeta_{0}={4\over 3}{m_{s}^{2}\over H^{2}}\log(2RH).
\end{equation}

To see the qualitative behavior of $\delta^{2}_{T}$, it would be
helpful to note ${\cal M}_{D}(\zeta_{0})\sim \zeta_{0}^{-{D+2\over 2}}$
when $\zeta_{0}\gg 1$. In this limit, we have
$\delta^{2}_{T}\sim ({H\over m_{pl}})^{4}(LH)^{D}(\log(2RH))^{-{D+2\over 2}}$
up to constant factors. $\delta^{2}_{T}$ is enhanced when
extra dimensions are large, $(LH)^{D}\gg 1$,  
since many fields contribute to it. $\delta^{2}_{T}$ becomes small
if $\log (2RH)$ were larger, since massive
fields weakly decay in time during inflation.

\paragraph*{Fixing parameters from the data.---}
We will now use observational data~\cite{WMAP},
\begin{equation}
 \delta_{\rm T}\sim 2.6\times 10^{-5}, \quad 
r_{t/s}\lesssim 0.22,
\label{WMAP}
\end{equation}
to constrain the parameters in our model. This implies
${H\over  m_{pl}} =\sqrt{\frac{9\pi}{2}
\delta_{\text{T}}^2r_{\text{t/s}}}
\lesssim 0.81\times 10^{-4}$.
Let us first assume this inequality is saturated.
Then the amplitude (\ref{amplitude}) provides the relation between
the two parameters $m_s$ and $L$, or equivalently,  
between $m_s$ and the string coupling $g_s$, since $L$ is written as 
$(Lm_s)^D=8\pi^6g_s^2(m_{pl}^2/m_s^2)$. 

Figs.~1 and 2 show $(Lm_{pl})$ and $g_{s}$ as functions of $m_{s}/H$, 
respectively. 
It is easier to have weak coupling 
with small $D$,  
while it is easier to keep $L$ not too large 
with large $D$.  
Typical values that are consistent with (\ref{WMAP})
would be: $\{D=2, m_{s}/H=0.2, Lm_{pl}=10^{12}, g_{s}=3\}$,
$\{D=3, m_{s}/H=0.2, Lm_{pl}=10^{10}, g_{s}=5\}$,
$\{D=4, m_{s}/H=0.1, Lm_{pl}=10^{9}, g_{s}=7\}$. The
number of the fields that participate in $\delta T/T$ is
roughly $N\sim (Lm_{s})^{D}$. For the above choice of parameters,
$10^{14}\lesssim N\lesssim 10^{16}$. The fact that $m_{s}$ is
close to $H$ (or slightly smaller) in our model can be understood
as follows: If we include quantum expectation value 
of the scalar fields on the r.h.s.\ of Friedmann equation during
inflation, we should have 
$H^2\gtrsim N H^{4}/m_{pl}^{2}$. Our values of $H$ and $N$
do not satisfy this.
This implies corrections (characterized by $H/m_{s}$) are 
so large that Friedmann equation is not applicable.

\begin{figure}[htb]
\begin{center}
\rotatebox{-90}{
\includegraphics[height=6cm]{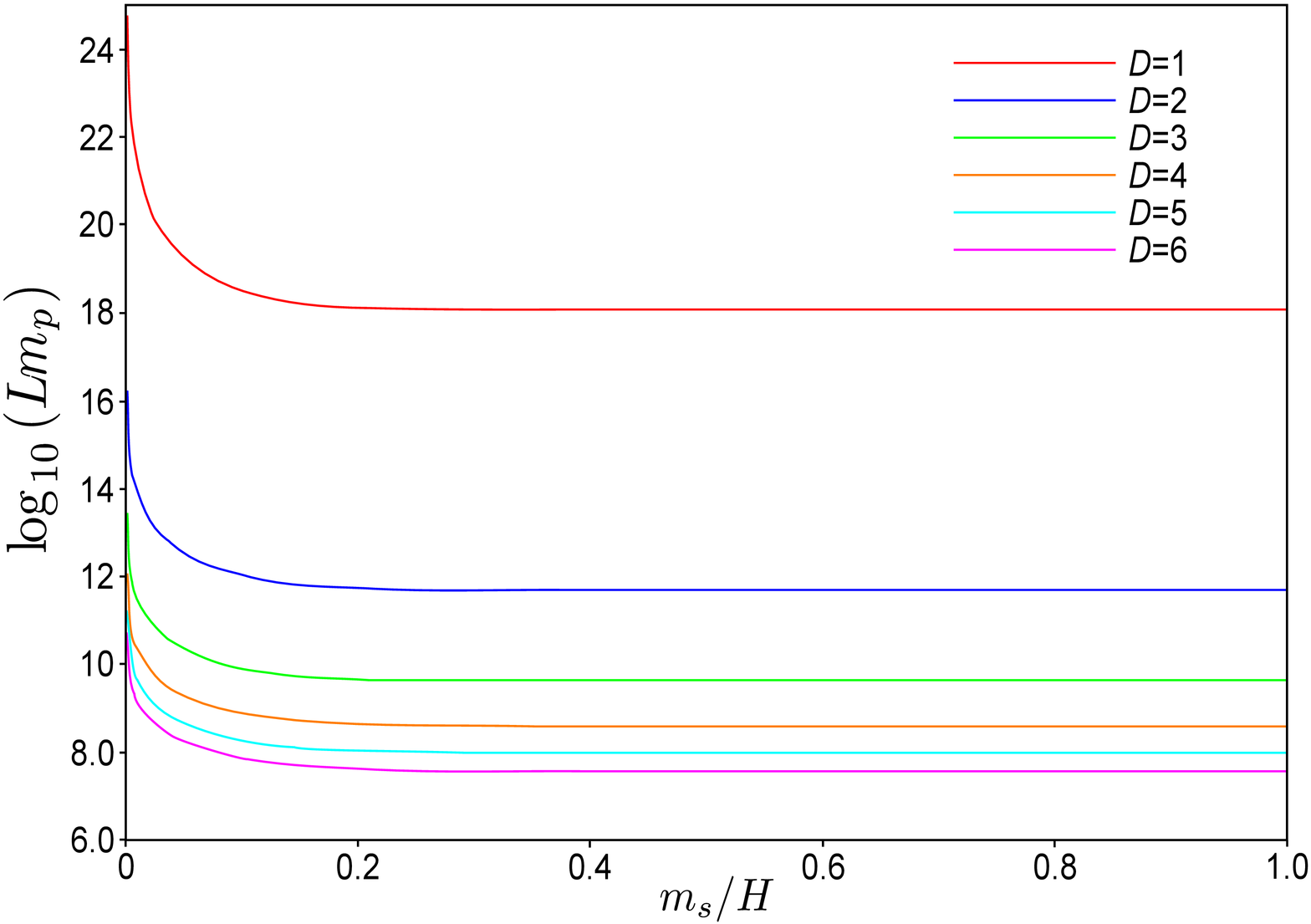}
}
\end{center}
\caption{$\log_{10}(Lm_{pl})$ as a function of 
$m_{s}/H$, with $\delta_{T}=2.6\times 10^{-5}$, 
$r_{\rm t/s}=0.22$, $RH\sim 10^{67}$. If $r_{\rm t/s}$ is smaller,
$(Lm_{pl})$ moves up; the $r_{\rm t/s}$ dependence is roughly 
$(Lm_{pl})\sim r_{\rm t/s}^{-({1/ 2}+{2/ D})}$.}
\label{Lms}
\end{figure}

\begin{figure}[htb]
\begin{center}
\rotatebox{-90}{
\includegraphics[height=6cm]{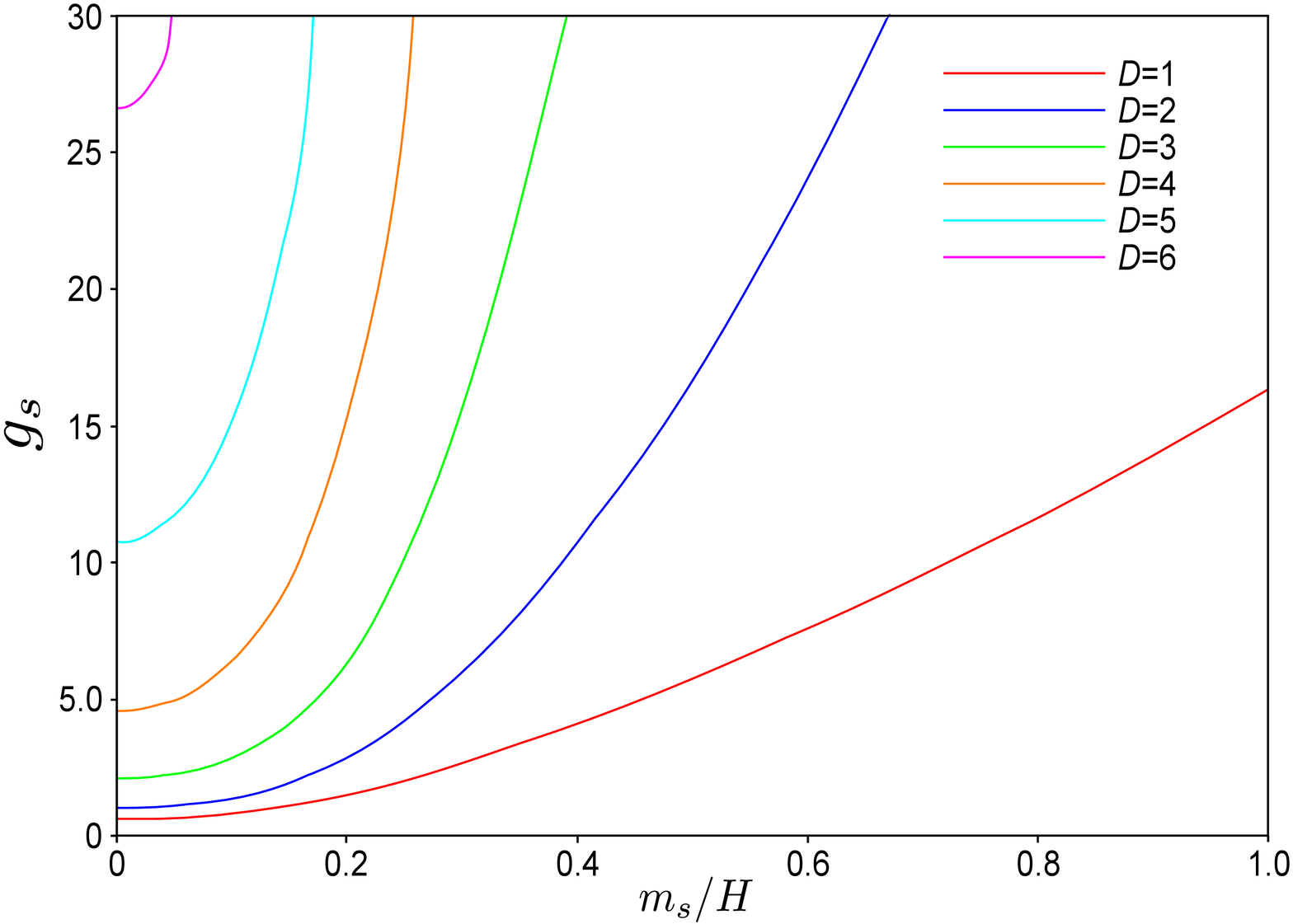}
}
\end{center}
\caption{$g_{s}$ as a function of $m_{s}/H$, with  
$\delta_{T}=2.6\times 10^{-5}$, 
$r_{\rm t/s}=0.22$, $RH\sim 10^{67}$. 
The dependence on $r_{\rm t/s}$ and $m_{s}/H$
at large $m_{s}/H$ is $g^{2}_{s}\sim ({m_{s}/ H})^{D+2}/r_{\rm t/s}$.}
\label{gsms}
\end{figure}

Since $\delta T/T$ originates from the fluctuations of
massive fields, the spectrum is slightly stronger in the
UV, thus the spectral index $n_s$ is larger than 1 in 
our model, 
\begin{eqnarray}
 n_{s}&=&1-{d\over d\log(Ha|\vec{x}-\vec{x'}|)}\log
\Big\langle{\delta T\over T}(\tau, \vec{x})
{\delta T\over T}(\tau, \vec{x'})\Big\rangle\nonumber\\
&=&1+{4m_{s}^{2}\over 3H^{2}}
{{\cal M}_{D}(\zeta_{0})\over {\cal M}_{D-2}(\zeta_{0})},
\end{eqnarray}
This is in the range $1\lesssim n_{s}\lesssim 1.02$
when $D=2$, and $1\lesssim n_{s}\lesssim 1.05$ for $D\le 6$. 

\paragraph*{Non-gaussianities.---}
The three-point function of $\Phi$ can readily be calculated.
It is given by the triangle diagram with each pair of 
points connected by $\langle\phi\phi\rangle$,
\begin{eqnarray}
&& \langle \Phi(\tau, \vec{x})\Phi(\tau, \vec{y})\Phi(\tau, \vec{z})\rangle
\\
&&\quad ={1\over 8\pi^{3}}\left({H\over m_{pl}}\right)^{6}\sum
\left(H^{3}a^{3}|\vec{x}-\vec{y}|
|\vec{y}-\vec{z}||\vec{x}-\vec{z}|\right)^{-\beta}.\nonumber
\end{eqnarray}
The non-linearity parameter $f_{\rm NL}$ is
defined by $\Phi\to \Phi_{g}+f_{\rm NL}\Phi_{g}^2$~\cite{Komatsu} 
with a gaussian field $\Phi_{g}$. Our three-point function
cannot be written this way with a local function $f_{\rm NL}$,
but we can estimate $f_{\rm NL}$ by considering
$\langle \Phi(x)\Phi(y)\Phi(z)\rangle$ at generic angular separation,
\begin{equation}
 f_{\rm NL}\sim {1\over 32}r_{\rm t/s}\left({m_{s}\over H}\right)^{2}
{{\cal M}_{D}({3\over 2}\zeta_{0})\over 
{\cal M}_{D-2}(\zeta_{0})}.
\end{equation}
This is proportional to $r_{\rm t/s}$, and further
suppressed.
(Note $\left({m_{s}\over H}\right)^{2}{\cal M}_{D}({3\over 2}\zeta_0)/
{\cal M}_{D-2}(\zeta_0)\sim (\log(2RH))^{-1}< 1$ when
${m_{s}\over H}\gg 1$.) For $r_{\rm t/s}=0.22$, we have
$f_{\rm NL}< 10^{-4}.$

\paragraph*{Conclusions.---}
We have studied the contribution to the CMB fluctuations
from quantum effects of many fields
with $m<H$. If Hubble is large, there could be many such 
fields. The effect studied in this Letter is important
when there are large extra dimensions. 

The simple model studied here gives $n_{s}>1$, 
but $n_{s}$ can be made smaller if $H$ decreases with time.
Time dependence of Hubble is also needed at the end of inflation.
We can study the dynamics of time dependent Hubble effectively
by considering an inflaton field $\varphi$.
By treating the inflaton fluctuation 
$\delta\varphi$ to be of the same order as $\Phi$, $\Psi$, we
obtain $\Phi$ in the super-horizon limit and in the slow-roll 
approximation,
\begin{eqnarray}
\Phi&\sim& -\left({V_{,\varphi}\over V}\right)^{2}
\left\{H_{*} \left({V\over V_{,\varphi}}\right)_{*}
+{1\over 4}\sum (\phi^{2}_{*}-\phi^{2})\right\}\nonumber\\
&&-\sum \pi G\beta \phi^{2} 
\end{eqnarray}
where the star denotes the quantities evaluated at the
horizon crossing, and $V$ is the inflaton potential. 
The first term is the part induced by the inflaton
fluctuation $\delta\varphi$, and the second term is the effect of
the matter $\phi_{A}$ that we have been studying.
The relative importance of the two terms depends on the details
of the model, such as the slope of the inflaton potential and the time
between horizon crossing and the end of inflation. The effect
of the large number of fields $\phi_{A}$ will be important
unless the slope is fine tuned to a 
small value.

More detailed analysis will be presented
in a forthcoming paper~\cite{fullpaper}.

\paragraph*{Acknowledgments.---}
We thank Toshihiro Matsuo for useful comments.  The work of
H.K is supported by the MEXT Grant-in-Aid for the Global COE 
Program, ``The Next Generation of Physics Spun from Universality and 
Emergence.'' 
H.K and M.N are supported by the JSPS Grant-in-Aid
for Scientific Research Nos. 18540264 and 21540290, respectively.
Y.S is supported by the MEXT
Grant-in-Aid for Young Scientists (B) No.~21740216.




\begin{thebibliography}{99}
\bibitem{WMAP} E.~Komatsu {\it et al.}  
  Astrophys.\ J.\ Suppl.\  {\bf 192}, 18 (2011).

\bibitem{HFKN}
M.~Fukuma, H.~Kawai and M.~Ninomiya, 
Int.\ J.\ Mod.\ Phys.\ A {\bf 19}, 4367 (2004); 
Y. Habara, H. Kawai and M. Ninomiya, 
Prog.\ Theor.\ Phys.\ Suppl.\ {\bf 164}, 7 (2007).

\bibitem{FSSY}
B.~Freivogel, Y.~Sekino, L.~Susskind and C.~P.~Yeh,
  Phys.\ Rev.\  D {\bf 74}, 086003 (2006);
Y.~Sekino and L.~Susskind,
  Phys.\ Rev.\  D {\bf 80}, 083531 (2009).

\bibitem{Nflation}S.~Dimopoulos,
S.~Kachru, J.~McGreevy and J.~G.~Wacker,
  JCAP {\bf 0808}, 003 (2008).

\bibitem{Liddle}
A.~R.~Liddle, A.~Mazumdar and F.~E.~Schunck,
  Phys.\ Rev.\  D {\bf 58}, 061301 (1998).

\bibitem{curvaton}
D.~H.~Lyth and D.~Wands,
  Phys.\ Lett.\  B {\bf 524}, 5 (2002).

\bibitem{vector}
  A.~Golovnev, V.~Mukhanov and V.~Vanchurin,
  JCAP {\bf 0806}, 009 (2008).

\bibitem{Mukhanov}
V.~Mukhanov, ``Physical Principles of Cosmology,''
Cambridge University Press, 2005. 


\bibitem{Komatsu}
  E.~Komatsu and D.~N.~Spergel,
  Phys.\ Rev.\ D {\bf 63}, 063002 (2001).

\bibitem{fullpaper}
Y.~Habara, H.~Kawai, M.~Ninomiya and Y.~Sekino, 
[arXiv:1110.5392 [hep-th]].


\end{thebibliography}
\end{document}